\documentstyle[preprint,revtex]{aps}
\begin{document}
\draft
\preprint{LU4899}
\begin{title}
Suppression of tunneling by interference\\
in half-integer--spin particles
\end{title}
\author{Daniel Loss\cite{AA}, David P. DiVincenzo, and G. Grinstein}
\begin{instit}
IBM Research Division\\
Thomas J. Watson Research Center\\
Yorktown Heights, New York 10598
\end{instit}
\receipt{July 27, 1992}
\begin{abstract}
Within a wide class of ferromagnetic and antiferromagnetic systems,
quantum tunneling of magnetization direction is spin-parity dependent:
it vanishes for magnetic particles with half-integer spin, but is
allowed for integer spin.  A coherent-state path integral calculation
shows that this topological effect results from interference between
tunneling paths.
\end{abstract}
\pacs{PACS: 75.10.Jm, 03.65.Sq, 75.30.Gw, 75.60.Jp}

\narrowtext
The search for new systems which exhibit quantum phenomena at the
mesoscopic scale has led to a great deal of activity in
recent years on nanometer-size magnetic particles.  In such particles
it is possible for the electronic spins to be locked together into a
well-ordered state, either aligned (ferromagnetic) or anti-aligned
(antiferromagnetic), but whose direction can rotate.  For
common forms of magnetic anisotropy the magnetic vector has two or
more low-energy directions.  Several recent investigations have
focussed on the possibility that the magnetic vector will pass between
these directions by quantum tunneling\cite{L,SCB}.  While the
tunneling rates are predicted theoretically to be exponentially small
in the size of the magnetic particles, recent experiments\cite{AS,BB}
suggest that quantum tunneling is observable in particles with
several thousand spins, which are available to present technology.

In this letter we show that for a wide range of systems, {\it quantum
tunneling is completely suppressed if the total spin of the magnetic
particle is half-integral but is allowed in integral-spin
particles}\cite{VH}.  An important experimental implication of this
result is that in ensembles of magnetic particles in which the exact
number of spins per particle is not precisely controlled (the typical
case with present technology), half of those particles (those with an
odd number of electrons) will not exhibit quantum tunneling.  Such
parity effects are well known in atomic physics, but have not been
previously noticed for magnetic particles\cite{DL}.  We show in
several specific examples below that this suppression has a
topological origin and arises as a destructive interference between
different tunneling paths.  Thus we find that one quantum effect
(tunneling) is suppressed by another (interference), leading to
``classical'' behavior in half-integer spin systems!

We begin with a re-analysis of the tunneling behavior of a small
ferromagnetic particle with easy-plane--easy-axis anisotropy in the
superparamagnetic limit where it behaves like a single large spin ---
referred to as ``Model I'' in \cite{CG}.  The classical anisotropy
energy $E$ has the form $E(\hat{\bf n})=E(\theta,\phi)=K_z\sigma^2
\cos^2\theta+K_y\sigma^2\sin^2\theta\sin^2\phi$, where $\hat{\bf n}$
is the magnetization direction and $K_z>K_y>0$; this energy
corresponds to a quantum spin Hamiltonian,
\begin{equation}
H=K_z\sigma_z^2+K_y\sigma_y^2, \label{QQ2}
\end{equation}
where $\sigma$ is the particle's total spin, and $M_0=\mu_B\sigma$ is
its magnetic moment, with $\mu_B$ the Bohr magneton.  We are
interested in the tunneling of the magnetization direction $\hat{\bf
n}$ between its two equivalent low-energy directions at the points
$\theta=\pi /2$, $\phi=0$, and $\theta=\pi /2$, $\phi=\pi$,
corresponding to the coherent-state kets $|0>$ and $|\pi >$,
resp.  To compute the tunneling rate $P$ \cite{R} we consider
the imaginary time transition amplitude expressed as a coherent-state
path integral for spins\cite{K,F,UnP,H},
\begin{equation}
<\pi|e^{-\beta H}|0>=\int_{\theta(0),\phi(0)}^{\theta(\beta),
\phi(\beta)}D\Omega e^{-S_E},\label{QQ4}
\end{equation}
where $\beta^{-1}$ is the temperature, $D\Omega\sim \prod _\tau
d\phi_\tau d\theta_\tau \sin\theta_\tau$, and where the integral runs
over all paths (i.e. magnetization directions) connecting one minimum
to the other.  The Lagrangean $L$ occurring in the Euclidean action
$S_E=\int_{0}^{\beta}d\tau L$ is given by
\begin{equation}
L=i\sigma\dot{\phi}-i\sigma\dot{\phi}\cos\theta+E\equiv
i\sigma\dot{\phi}+L_0.\label{QQ6}
\end{equation}
The first two terms of Eq. (\ref{QQ6}) define the Wess-Zumino
term\cite{F} which is of crucial importance in the following.  For any
path on the 2-sphere $S_2$, parameterized by $\phi(\tau)$ and
$\theta(\tau)$, this contribution to the action is equal to $i\sigma$
times the area swept out on $S_2$ between the path and the north pole;
for closed paths this has exactly the form of the Berry
phase\cite{H,B,LG}.

The first term of Eq. (\ref{QQ6}) has some special features which
require discussion.  It is a total derivative term, which,
when integrated, gives the boundary contribution
\begin{equation}
i\sigma\int_{0}^{\beta}d\tau\dot{\phi}=i\sigma(\phi(\beta)-\phi(0)
+2n\pi).\label{QQ7}
\end{equation}
to $S_E$.  Here $n$ is a winding number counting the number of times
which the path wraps around the north pole.  As a total derivative,
this term has no effect on the classical (Bloch) equations of motion,
which can be derived by extremizing just the $L_0$ piece of the
action\cite{CG}.  In consequence, this boundary term
is commonly ignored.  We show
here, however, that this term
is crucial for the quantum properties of the
magnetic particle, making the tunneling behavior of integral and
half-integral spins strikingly different.

This result is most clearly seen by treating the transition amplitude
of Eq. (\ref{QQ4}) within an instanton approximation (i.e.,
saddle-point evaluation)\cite{R}.  In the easy-plane--easy-axis model,
the saddle point paths remain near $\theta=\pi/2$ (cf. \cite{foof}),
so they can be characterized by their winding in the $\phi$ variable
only.  This model has the important feature that the passage from
$\phi=0$ to $\phi=\pi$ can be accomplished either by an instanton or
an anti-instanton path, i.e., a clockwise or counterclockwise winding
over the barrier; see Fig. \ref{F1}(a).  Then the propagator of Eq.
(\ref{QQ4}) is approximated by a sum over paths comprising a sequence
of instantons and anti-instantons winding over the barrier.  The
expression for low temperatures is
\begin{eqnarray}
<\pi|e^{-\beta H}|0>\propto e^{-\beta
E_0} \sum_{m,l\ge 0}^{m+l\: odd}\frac{(D\beta)^{l+m}}{m!\: l!}
e^{i\sigma\pi(m-l)}\nonumber\\
\times e^{-S_0^{cl}(m+l)}=e^{-\beta
E_0}\sinh[2D\beta\cos(\pi\sigma)e^{-S_0^{cl}}].\label{QQ8}
\end{eqnarray}
Here $D$ is the fluctuation-determinant (without zero-mode)\cite{R},
$l$ is the number of instantons and $m$ the number of anti-instantons
in the path, $E_0$ is the zero-point energy in one well,
$\exp(-S_0^{cl})$ is the instanton contribution to the path integral,
and the constraint on the sum reflects the requirement that
$\phi(\beta)=\pi (mod(2\pi))$.  Note that
the action $S_0^{cl}$ is exactly the
same for the instanton and anti-instanton, because
this action is unchanged if $\phi$ is replaced by
$-\phi$\cite{foof}.
{}From Eq. (\ref{QQ8}) we can now
read off the tunneling rate $P$ (energy level splitting):
\begin{equation}
P=4D|cos(\pi\sigma)|e^{-S_0^{cl}},\:\:e^{-S_0^{cl}}=
\left(\frac{1-\sqrt{\lambda}}{1+\sqrt{\lambda}}\right)^\sigma.
\label{QQ8a}
\end{equation}
Here $\lambda\equiv K_y/K_z$.  Evidently, the tunneling rate vanishes
for half-integral spins because of the $\cos(\pi\sigma)$-factor which
arises directly from the topological boundary term of Eq.
(\ref{QQ7}).  This factor represents an interference between the
instanton and anti-instanton contributions to tunneling.  If the spin
$\sigma$ is an integer, then the interference is constructive, and the
total tunneling rate is of order of the single-instanton rate.  But if
the spin is half-integral, then $\cos(\pi\sigma)=0$; there is {\it
destructive interference} between the instanton and anti-instanton,
and the tunneling rate is {\it zero}.  Note that this spin-parity
effect is of topological origin and thus independent of the magnitude
of the spin.

A graphic illustration of the difference between integer and
half-integer spin systems is afforded by the numerical energy spectra
of Hamiltonian (\ref{QQ2}) presented in Fig. \ref{F2}. Results for
$\sigma=10\frac{1}{2},\: 10$ are shown as a function of $K_y$, for
$K_z=1$. As $K_y$ is increased, a barrier between the two easy-axis
directions is developed, and tunneling occurs for the integer spin
case.  The tunneling rate, which is proportional to the energy
splitting between the two lowest energy levels, decreases with
increasing $K_y$ (see Eq. (\ref{QQ8a})), eventually vanishing when
$K_y$ = $K_z$, since at this point the $x$-component of the
magnetization commutes with the Hamiltonian and so is conserved.  In
the half-integer spin case, by contrast, it is easy to show from
direct consideration of Hamiltonian (\ref{QQ2}) that the space of
states decomposes into two independent subspaces with identical energy
spectra.  Thus all states are strictly doubly degenerate, and there is
no tunneling, consistent with the arguments above.

The result that tunneling is suppressed for half-integer spin has much
greater generality than is suggested by the above analysis.  The
suppression can be derived within the coherent-state path integral
formalism independent of any approximation.  Any arbitrary path
$\theta(\tau)$, $\phi(\tau)$ in Eq. (\ref{QQ4}) (not just the
saddle-point path) can always be paired with another path,
$\pi-\theta(\tau)$, $-\phi(\tau)$, which has the same $L_0$ in Eq.
(\ref{QQ6}), while the winding-number term of Eq. (\ref{QQ6}) is
reversed; thus, the destructive interference for half-integral spin
occurs term-by-term in Eq. (\ref{QQ4}).  Furthermore, this pairing is
possible for much more general Hamiltonians than Eq. (\ref{QQ2}).
One generalization involves adding any additional terms to $H$ which
preserve the symmetry of rotation around the x-axis (e.g.,
$\sigma_z\sigma_y$), including odd-order terms like $B_{ext}\sigma_x$.
Of course, if such an external-field term is present, then the
degeneracy between the two wells is broken; nevertheless, the
vanishing of the propagator of Eq. (\ref{QQ4}) for half-integer spin
still implies vanishing hybridization between the two (inequivalent)
wells, and an absence of tunneling.  The other interesting
generalization of $H$ involves adding {\it any} even-order term (e.g.,
$\sigma_z^6$, $\sigma_x\sigma_y\sigma_z^2$), but excluding odd-order
terms.  Then H has time reversal (T) invariance; absence of tunneling
(i.e., vanishing of the transition amplitude (\ref{QQ8})) can again be
proved by a pairing of paths (this time, the paths $\hat{\bf
n}(\tau),\:-\hat{\bf n}(\beta-\tau)$).  In this case, the vanishing of
tunnel splitting is directly related to the Kramers
degeneracy\cite{M}.  We can say more generally that for anisotropy
Hamiltonians which have two equivalent energy minima, the absence of
tunneling for half-integral spin follows directly from the Kramers
theorem\cite{foof2}.

We now analyze a problem raised by a recent experiment on
tunneling in small {\it antiferromagnetic} particles\cite{AS}.
The particles were not perfect antiferromagnets, but carried an excess
magnetic moment, presumed to arise from surface effects, of a few
percent of the total number of spins.  In applying to the experiment
the theory developed in Ref. \cite{BC} for the tunneling of a
N\'eel-ordered antiferromagnetic particle through an anisotropy
barrier, it was assumed \cite{AS} (as suggested in \cite{BC}), that
the excess spins simply follow adiabatically the direction of the
N\'eel vector without affecting the tunneling dynamics.  We show now
that, within a simple extension of the existing models, this
expectation is false:  As above, a {\it half-integer} excess spin of
any size quenches tunneling; an integer excess spin also influences
the tunneling rate, though more modestly.

Our model is a 1D antiferromagnetic ring with $N$ spins
$\mbox{\boldmath$\sigma$}^{(j)}$ (spin magnitude $\sigma$) and
periodic boundary conditions; all the spins $\mbox{\boldmath$\sigma$}$
are coupled to a central excess spin $\bf{s}$ with coupling constant
$J_c^{(j)}\equiv(-1)^jJ_c$.  Thus the central spin will prefer to remain
aligned with the N\'eel vector of the spins on the ring (see Fig.
\ref{F1}(b)).  The Hamiltonian of the system may be written
\begin{equation}
{\cal{H}}=\sum_{j=1}^{N}
[J\mbox{\boldmath$\sigma$}^{(j)}\!\cdot\!
\mbox{\boldmath$\sigma$}^{(j+1)}
+H(\mbox{\boldmath$\sigma$}^{(j)})
+J_c^{(j)}\mbox{\boldmath$\sigma$}^{(j)}\!\cdot\!\bf{s}].\label{QQ14}
\end{equation}
Here $N$ is even; odd $N$ would produce a different problem involving
frustration and non-zero total spin on the ring.  $H$ is the
anisotropy Hamiltonian (\ref{QQ2}), and leads to two degenerate
states, the one shown in Fig. \ref{F1}(b), and the one where all the
spin directions are reversed.  The problem is to compute the tunneling
rate between these two states.

This model may be considerably simplified if we restrict our
consideration to the continuum and semi-classical (large $\sigma$)
limit.  For this we use again the coherent-state path integral
representation, and, applying standard manipulations\cite{F,H}, we
then arrive at a generalized non-linear $\sigma$-model in terms of a
N\'eel unit-vector $\hat{\bf{l}}$, and a central spin unit vector
$\hat{\bf{n}}_s$\cite{footnote2}.  We take $\hat{\bf{l}}$ to be
uniform around the ring, obtaining the following effective Euclidean
action:
\begin{eqnarray}
S_E=\int_{0}^{\beta} d\tau[\frac{\chi_\perp}{8\mu_B^2}
(\dot{\theta}_l^2+\dot{\phi}
_l^2\sin^2\theta_l)+NE(\theta_l,\phi_l)\nonumber\\
+\sigma N J_c\hat{\bf{l}}
\cdot\hat{\bf{n}}_s+is\dot{\phi}_s(1-\cos\theta_s)],\label{QQ15}
\end{eqnarray}
where $\hat{\bf{l}}$ and $\hat{\bf{n}}$ are expressed in polar
coordinates, and the transverse susceptibility $\chi_\perp$ is related
to the parameters of Eq. (\ref{QQ14}) by $\chi_\perp=N\mu_B^2/J$.

As before, we find that if the excess spin $s$ of the antiferromagnet
is half-integer, then the tunneling rate is exactly zero.  The proof
follows in the same way:  we consider an arbitrary path
$\{\theta_l(\tau),\phi_l(\tau), \theta_s(\tau),\phi_s(\tau)\}$ and its
partner with $\theta_{l,s} \rightarrow\pi-\theta_{l,s}$,
$\phi_{l,s}\rightarrow-\phi_{l,s}$.  Then in the path integral
expression analogous to Eq. (\ref{QQ4}), and with $\hat{\bf{l}}$ and
$\hat{\bf{n}}_s$ having the same boundary conditions, these two paths
have opposite winding-number contributions $\pm is\pi(1+2n)$ (cf. Eq.
(\ref{QQ6})), and the same values of $L_0$.  Thus, a factor $\cos(\pi
s)$ appears exactly as in Eq. (\ref{QQ8}), implying complete
destructive interference and a vanishing of the tunneling rate if the
central spin $s$ is half-integral.

Also as in the ferromagnetic case, this vanishing can be seen to be
related to the Kramers degeneracy.  Again, the model Eq.
(\ref{QQ14}), and therefore Eq. (\ref{QQ15}), has time reversal
invariance (since all terms contain an even number of spin operators).
Thus, the ground state is a Kramers doublet so long as the total spin
of the model is half-integral, which, since $N$ is even, requires that
$s$ is half-integral.  Again, suppression of tunneling is related to
the absence of a tunnel splitting in the ground state.  However, we
caution as we did above that there is not a one-to-one correspondence
between the Kramers theorem and absence of tunneling for
antiferromagnetic models.

Finally, we consider the question of how strongly a nonzero but {\it
integer} $s$ modifies the tunneling dynamics of Eq. (\ref{QQ15}).
For $s=0$ this model can be solved in the instanton approximation; the
saddle point solution happens to be identical to a different
(uniaxial-anisotropy) model considered in \cite{BC}.  The saddle point
path has $\theta_l=\pi/2$ everywhere, while $\phi_l$ passes from 0 to
$\pi$.  As in \cite{BC}, the tunneling rate is given by:  $P\sim\omega
e^{-\sqrt{2\chi_\perp k_y}/\mu_B}$, with
$\omega\equiv\mu_B(8k_y/\chi_\perp)^{\frac{1}{2}}$, and with the
notation $k_{y,z}\equiv{\sigma}^2 NK_{y,z}$ for the anisotropy
constants of the whole ring.

A full solution of Eq. (\ref{QQ15}) for $s\ne 0$ seems difficult.  We
can make progress in the adiabatic approximation\cite{B,LG}, in which
the spin ${\hat{\bf n}}_s$ simply follows the instantaneous direction
of the N\'eel vector $\hat{\bf l}$ at every point along the path.
In Eq. (\ref{QQ15}), this simply involves removing the $\hat{\bf
l}\cdot{\hat{\bf n}}_s$ term (since it is just a constant) and setting
$\theta_s=\theta_l$, $\phi_s=\phi_l$\cite{footnote4}.

We can now find an approximate solution for the tunneling rate if we
assume that $k_y\ll k_z$ and that therefore $\theta_l$ does not
fluctuate very far away from $\pi/2$.  If we write
$\theta_l=\pi/2+\vartheta$ and expand the Lagrangean to second order
in $\vartheta$, we obtain
\begin{eqnarray}
L\approx\frac{\chi_\perp}{8\mu_B^2}(\dot{\vartheta}^2+\dot{\phi}^2)
+{\vartheta}^2(k_z-k_y\sin^2 \phi - \frac{\chi_\perp}{8\mu_B^2}
\dot{\phi}^2)\nonumber\\
+k_y\sin^2\phi+is\dot{\phi}(1+\vartheta).\label{QQ19}
\end{eqnarray}
We find\cite{footnote5} that for $k_y\ll k_z$ we have
$\dot{\vartheta}^2\ll \dot{\phi}^2\ll 8\mu_B^2 k_z/{\chi_\perp}$.
Keeping then only the leading terms in $L$ we find for the extremal
$\vartheta$-path, $\vartheta\approx -is\dot{\phi}/2k_z$, obtaining an
effective Lagrangean for the $\phi$ variable,
$L\approx(\chi_\perp/8\mu_B^2+s^2/4k_z)\dot{\phi}^2 +k_y\sin^2\phi +
is\dot{\phi}$.  This Lagrangean is now identical to the one for a pure
antiferromagnet (see \cite{BC}) with $s=0$, but with a topological
boundary term and a modified value of the transverse susceptibility
(the ``moment of inertia of the rigid rotor''\cite{UnP}):
$\chi_\perp^{eff}=\chi_\perp+2s^2\mu_B^2/k_z$.  Thus the tunneling
rate is finally given by
\begin{equation}
P\sim |cos(\pi s)|\omega e^{-\sqrt{2\chi_\perp^{eff}k_y}/\mu_B}.
\label{QQ20}
\end{equation}
Again, for half-integral excess spin $s$ tunneling is suppressed.  For
integral $s$ we can define a crossover excess spin $s_c$ for which the
contribution of the excess spin to the moment of inertia becomes
comparable to that of the antiferromagnetic ring:
$s_c=\sqrt{\chi_\perp k_z/2}/\mu_B$.  If, e.g., $k_z\approx 10k_y$,
then $s_c$ is approximately 1.6 times the WKB exponent of Eq.
(\ref{QQ20}) for $s=0$. As Ref. \cite{BC} has pointed out, for
practical reasons this WKB exponent cannot be much larger than about
25 if tunneling is to be observed in a small particle.  Thus, the
magnitude of the excess spin is quite restricted ($s<1.6\times 25$)
if pure antiferromagnetic dynamics is to be observed.

In summary, we have demonstrated that tunneling in a wide class of
magnetic particles is strongly parity dependent, being completely
suppressed for half-integral spins.  Preliminary results indicate
that similar effects are to be expected in the tunneling of domain
walls\cite{SCB}.  We expect that these phenomena can still exist
in the presence of moderate dissipation\cite{UnP}.

\figure{(a) Anisotropy energy $vs.$ $\phi$
at $\theta=\pi/2$, showing the path for the instanton ({\it I}), the
anti-instanton ({\it AI}), and a more general path $\cal{P}$
containing one
anti-instanton and four instantons.  (b) Antiferromagnetic ring
coupled to an excess spin (Eq. (\ref{QQ14})).\label{F1}}
\figure{Low-lying eigenenergies as a function of $K_y$ for $K_z=1$
for the model of Eq. (\ref{QQ2}) for
spin-$10\frac{1}{2}$ and spin-10.  All levels are doubly-degenerate
for $\sigma=10\frac{1}{2}$.
\label{F2}}

\begin{references}
\bibitem[*]{AA} Permanent address as of January 1993: Physics Dept.,
Simon Fraser University, Burnaby, British Columbia V5A 1S6.
\bibitem{L} A. J. Leggett {\it et al.}, Rev. Mod. Phys. {\bf 59}, 1
(1987).
\bibitem{SCB} For a review with references, see
P. C. E. Stamp, E. M. Chudnovsky, and B. Barbara,
Int. J. Mod.  Phys. B {\bf 6}, 1355 (1992).
\bibitem{AS} D. D. Awschalom, J. F. Smyth, G. Grinstein, D. P.
DiVincenzo, and D. Loss, Phys. Rev. Lett. {\bf 68}, 3092 (1992).
\bibitem{BB} C. Paulsen {\it et al.}, Phys. Lett. {\bf A161}, 319
(1991).
\bibitem{VH} As discussed later, under certain conditions this result
follows from the Kramers degeneracy, a fact first noted by J. L. Van
Hemmen and S. S\"uto, Europhys. Lett. {\bf 1}, 481 (1986);
Physica {\bf 141B}, 37 (1986).
\bibitem{DL} However, see D. Loss, Phys. Rev. Lett. {\bf 69}, 343 (1992).
\bibitem{CG} E. M. Chudnovsky and L. Gunther, Phys. Rev. Lett. {\bf 60},
661 (1988).
\bibitem{R} R. Rajaraman, {\it Solitons and Instantons} (North Holland,
Netherlands, 1987).
\bibitem{K} J. Klauder, Phys. Rev. D {\bf 19}, 2349 (1979).
\bibitem{F} E. Fradkin, {\it Field Theories in Condensed Matter}
(Addison-Wesley, Reading, MA, 1991), Chap. 5.
\bibitem{UnP}
Further details of our analysis will appear in a future
publication:  D. Loss, D. P. DiVincenzo, and G. Grinstein
(unpublished).
\bibitem{H} F. D. M. Haldane, Phys. Rev. Lett. {\bf 57}, 1488 (1986);
{\bf 61}, 1029 (1988).
\bibitem{B} M. V. Berry, Proc. R. Soc. London, {\bf 392A}, 45 (1984).
\bibitem{LG} D. Loss and P. Goldbart, Phys. Rev. B {\bf 45},
13544 (1992).
\bibitem{foof} The equations of motion obtained from Eq. (\ref{QQ6})
(cf. \cite{CG}) can be put into the form:
$\sin^2\theta=(1-\lambda\sin^2\phi)^{-1}$, and
$\dot{\phi}^2=\omega_0^2(1-\lambda\sin^2\phi)\sin^2\phi$.  Here
$\omega_0\equiv2\sigma (K_yK_z)^{\frac{1}{2}}$.  Since $\lambda < 1$,
$\sin^2\theta > 1$, and $\theta$ must be complex (for finite-action
solutions $\phi$ has to be real); in particular, it can be written in
the form $\theta=\pi/2+ i\theta_I$ (with $\theta_I$ real).  This is
the sense in which the instanton path remains ``close to''
$\theta=\pi/2$.  As a consequence of this complexification the
anisotropy energy $E$ evaluated along the classical path {\it
vanishes}, and $L_0^{cl}=2K_y\sin^2\phi^{cl}$ with $\phi^{cl}=\pm
\arccos(\sqrt{1-\lambda}\tanh(\omega_0\tau)/\sqrt{1-\lambda\tanh^2(
\omega_0\tau}))$.
\bibitem{M} A. Messiah, {\it Quantum Mechanics, vol. II} (Wiley, New
York, 1962), p. 675, 750ff.
\bibitem{foof2} The presence of Kramers degeneracy does not rule out
tunneling in all cases.  For a nearly-cubic anisotropy
$H=K(\sigma_x^4+\sigma_y^4+\sigma_z^4)+\epsilon \sigma_x^2$ a
tunnel-splitting is developed between pairs of doubly degenerate
states.  This is a consequence of the more complex multi-well
structure of this potential when $\epsilon=0$; tunneling can take
place between any (111) directions, not just between antipodal points
as assumed in Eq. (\ref{QQ4}).
\bibitem{BC} B. Barbara and E. M. Chudnovsky, Phys. Lett. A{\bf145},
205 (1990).
\bibitem{footnote2} Since typically $K_{y,z}/J \ll 1$, we drop terms
involving this small parameter\cite{UnP}.
\bibitem{footnote4} The adiabatic approximation can be justified by an
extension of standard techniques\cite{B,LG,M}; the condition for its
validity is that the Zeeman energy $N\sigma J_c$ be greater than the
characteristic tunneling attempt rate $\hbar\omega$.
Note that the
remaining $2s$ terms occurring in the adiabatic approximation have
higher energy and can be dropped at low temperatures.
\bibitem{footnote5}
The approximations adopted after Eq. (\ref{QQ19}) can be justified as
follows:  The characteristic tunneling time is given by $\omega^{-1}$
(see Eq. (\ref{QQ20})).  Thus $\dot{\phi}\sim\omega$, $\ddot\phi
\sim\omega^2$, and therefore $\dot{\vartheta}\sim s \omega^2/k_z$.
Thus the condition that $\dot{\vartheta}$ can be ignored compared with
$\dot\phi$ is $s\ll s_c
\sqrt{k_z/k_y}/2=k_z\sqrt{2k_y\chi_\perp}/4\mu_B k_y$.
Similarly, we also have $\dot{\phi}^2\sim \omega^2\ll
8\mu_B^2k_z/{\chi_\perp}$, if $k_y\ll k_z$.
\end{references}
\end{document}